\documentclass[twocolumn,twoside,slac_two]{revtex4}
\usepackage{subfigure}
\usepackage{graphicx}
\usepackage{fancyhdr}
\def\apj{ApJ}  
\def\aap{A\&A} 

\begin{document}

\title{On the phenomenological classification of continuum radio spectra
  variability patterns of {\it Fermi} blazars}

%

\author{E. Angelakis, L. Fuhrmann, I. Nestoras, C. M. Fromm, R. Schmidt, J. A. Zensus, N. Marchili, T. P. Krichbaum}
\affiliation{Max-Planck-Institut f\"ur Radioastronomie, Auf dem H\"ugel 69, DE-53121,  Bonn, Germany}
\author{M. Perucho-Pla}
\affiliation{Department d'Astronomia i Astrof\'{i}sica, Universitat de Val\`{e}ncia, C/Dr. Moliner 50, 46100 Burjassot, Val\`{e}ncia, Spain}
\author{H. Ungerechts, A. Sievers, D. Riquelme}
\affiliation{Instituto de Radio Astronomía Milim\'{e}trica, Avenida Divina Pastora 7, Local 20 E 18012, Granada, Spain}

\begin{abstract}
  The {\em F-GAMMA} program is a coordinated effort to investigate the physics of Active
  Galactic Nuclei (AGNs) via multi-frequency monitoring of {\em Fermi} blazars. The
  current study is concerned with the broad-band radio spectra composed of measurement at
  ten frequencies between 2.64 and 142\,GHz. It is shown that any of the 78 sources
  studied can be classified in terms of their variability characteristics in merely 5
  types of variability. The first four types are dominated by spectral evolution and can
  be reproduced by a simple two-component system made of the quiescent spectrum of a large
  scale jet populated with a flaring event evolving according to
  \cite{Marscher1985ApJ}. The last type is characterized by an achromatic change of the
  broad-band spectrum which must be attributed to a completely different mechanism. Here
  are presented, the classification, the assumed physical system and the results of
  simulations that have been conducted.
\end{abstract}

\maketitle

\thispagestyle{fancy}

\section{INTRODUCTION}
Among the most evident characteristics of blazars is the intense variability at all
wavelengths. Studies of the variability characteristics, preferably with simultaneous
data, can shed light on the physics driving the energy production and dissipation in these
systems \citep[e.g.][]{boettcher2010,boettcher2010HEAD}. The {\em F-GAMMA} program
\citep[see][]{fuhrmann2007AIPC,angelakis2008MmSAI..79.1042A,2010arXiv1006.5610A} is a coordinated
effort to explore exactly this very possibility by monthly monitoring of {\em Fermi}
blazars. {\em F-GAMMA} is covering mostly the radio cm to sub-mm bands primarily
with the Effelsberg 100-m, the IRAM 30-m and the APEX 12-m telescopes (although optical
telescopes are participating as well, Fuhrmann et al. in prep.) for roughly 60 prominent
blazars.

The cause for the variability itself has been long debated. The ``shock-in-Jet'' model
suggested by \cite{Marscher1985ApJ}, is the most accepted one and attributes the
variability to shocks propagating down the jet. The basic assumption is that changes at
the onset of the jet, (e.g. changes in the injection rate, the magnetic field, bulk
Lorentz factor etc.) cause the formation of shocks, which then suffer first {\em Compton},
then {\em synchrotron} and finally {\em Adiabatic} losses. This is the main model
prediction which is used in the following.

Here it is argued that (a) the variability patterns can be classified in only 5
phenomenological types and (b) the different phenomenological classes can be reproduced
with a simple system composed of a quiescent spectrum populated by a flaring event evolving
according to the ``shock-in-Jet'' model.

\section{OBSERVATIONS AND DATA REDUCTION}

\begin{figure*}
  \centering
  \subfigure[Type 1]{\includegraphics[width=70pt,height=150pt,angle=-90]{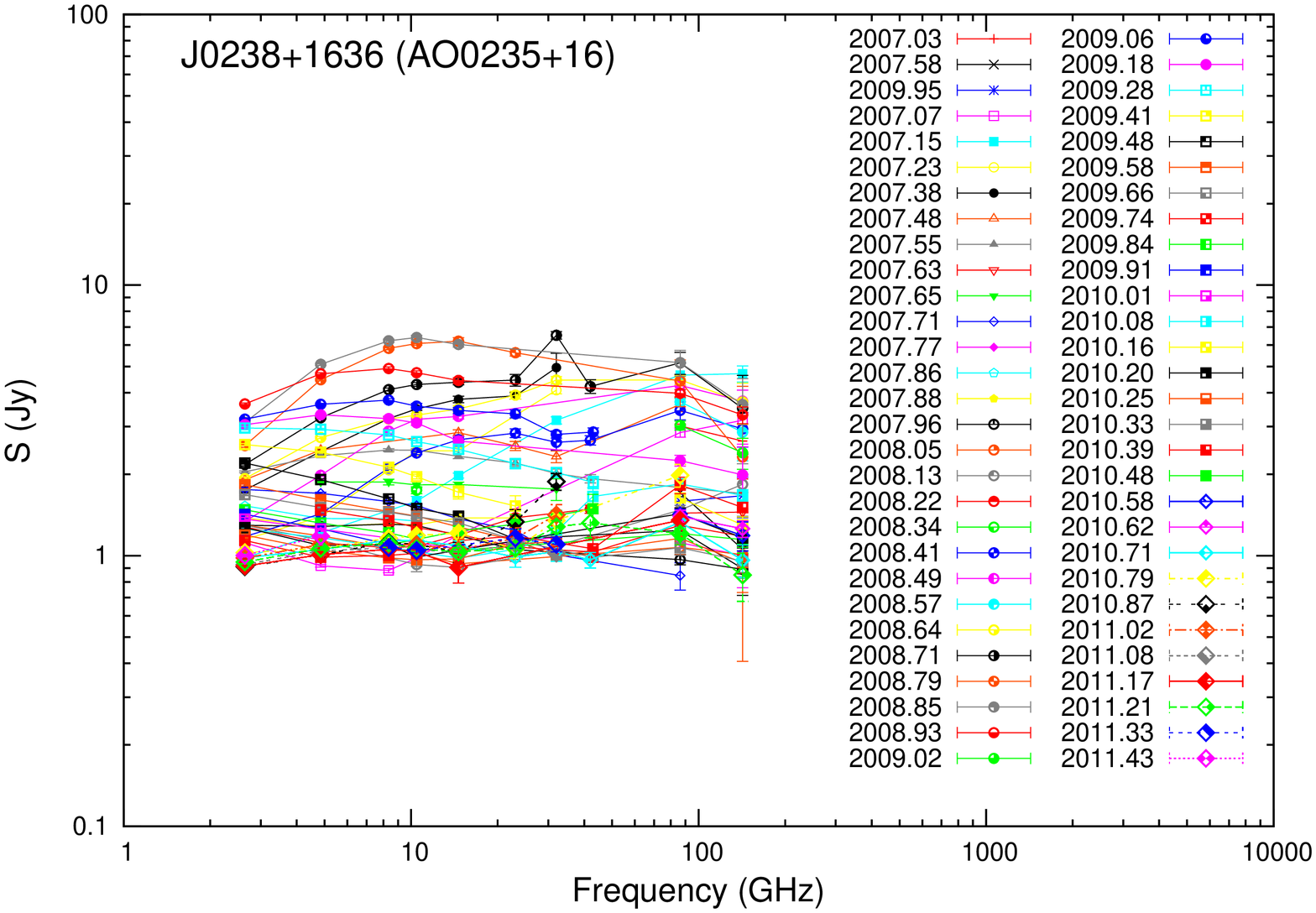} \label{fig:t1}}  
  \subfigure[Type 1b]{\includegraphics[width=70pt,height=150pt,angle=-90]{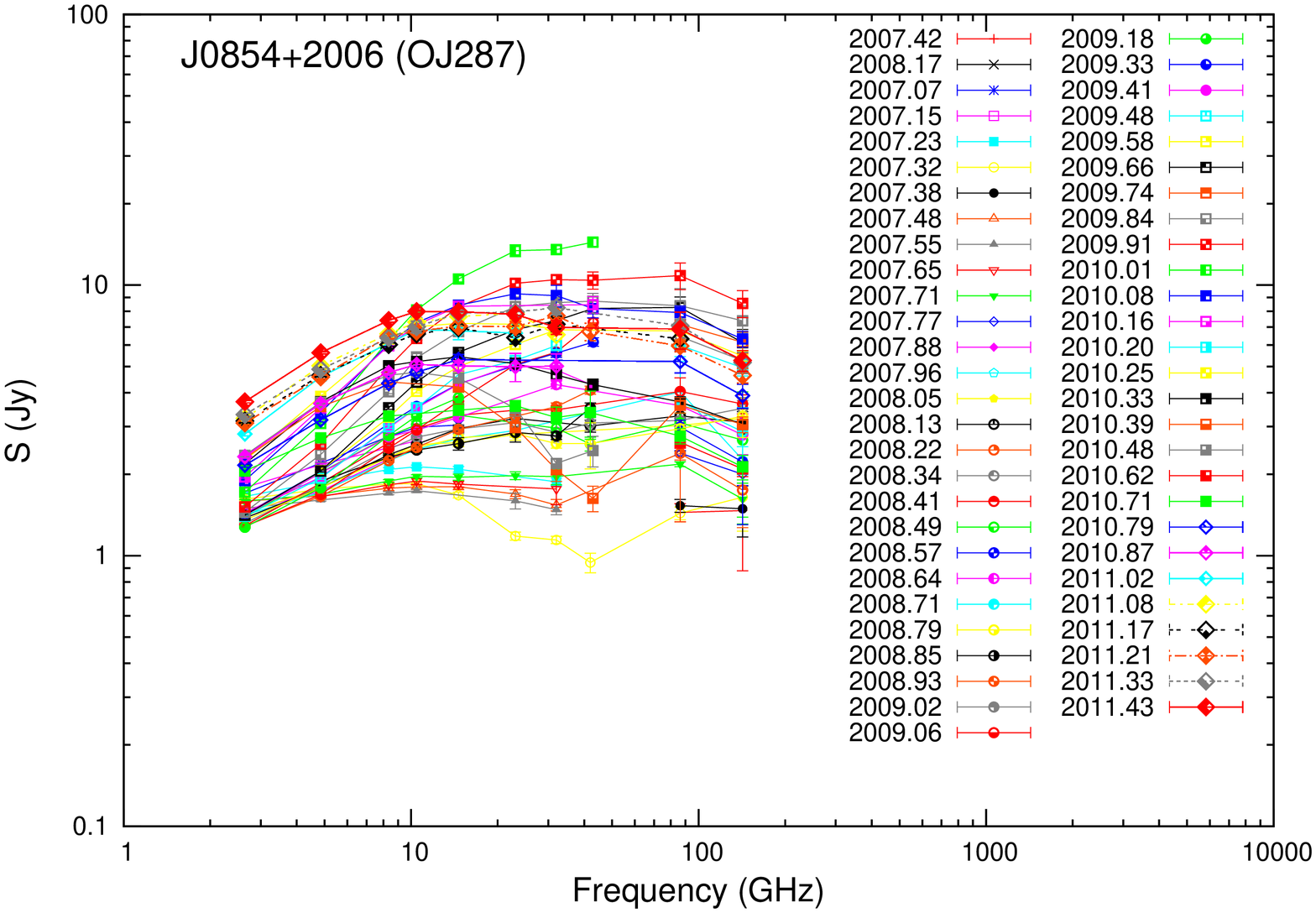} \label{fig:t1b}}
  \subfigure[Type 2]{\includegraphics[width=70pt,height=150pt,angle=-90]{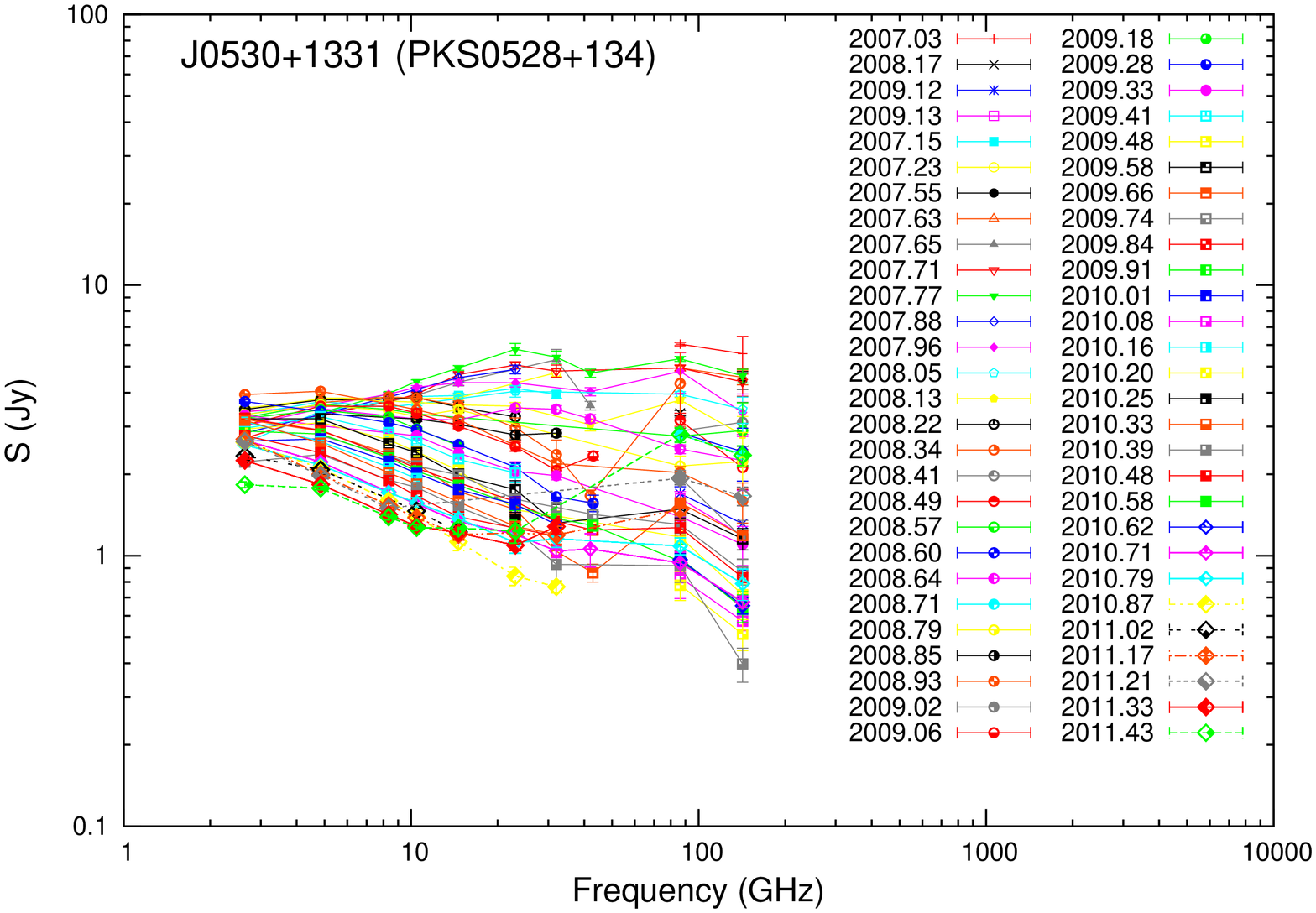} \label{fig:t2}}
  \\
  \subfigure[Type 3]{\includegraphics[width=70pt,height=150pt,angle=-90]{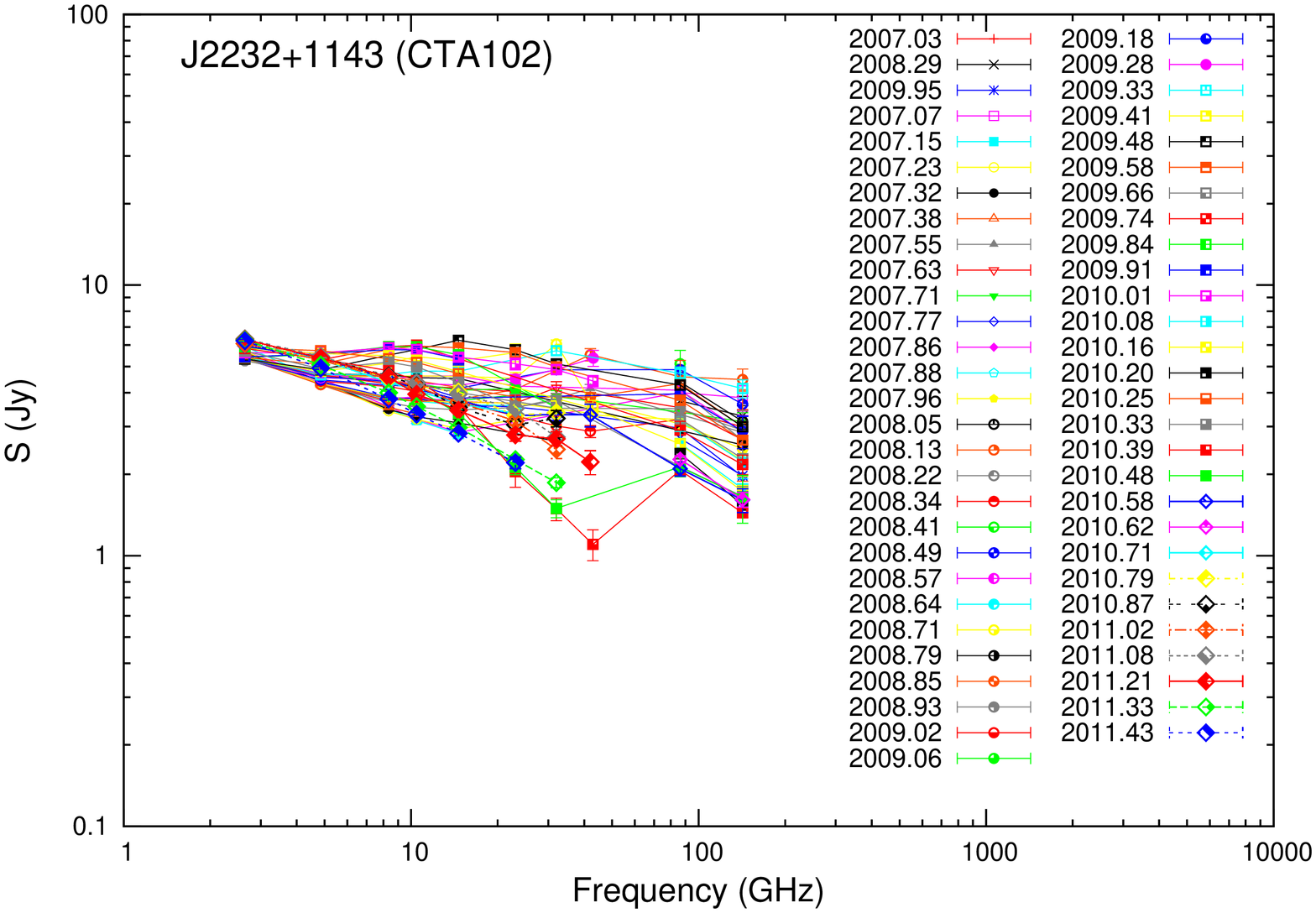} \label{fig:t3}}  
  \subfigure[Type 3b]{\includegraphics[width=70pt,height=150pt,angle=-90]{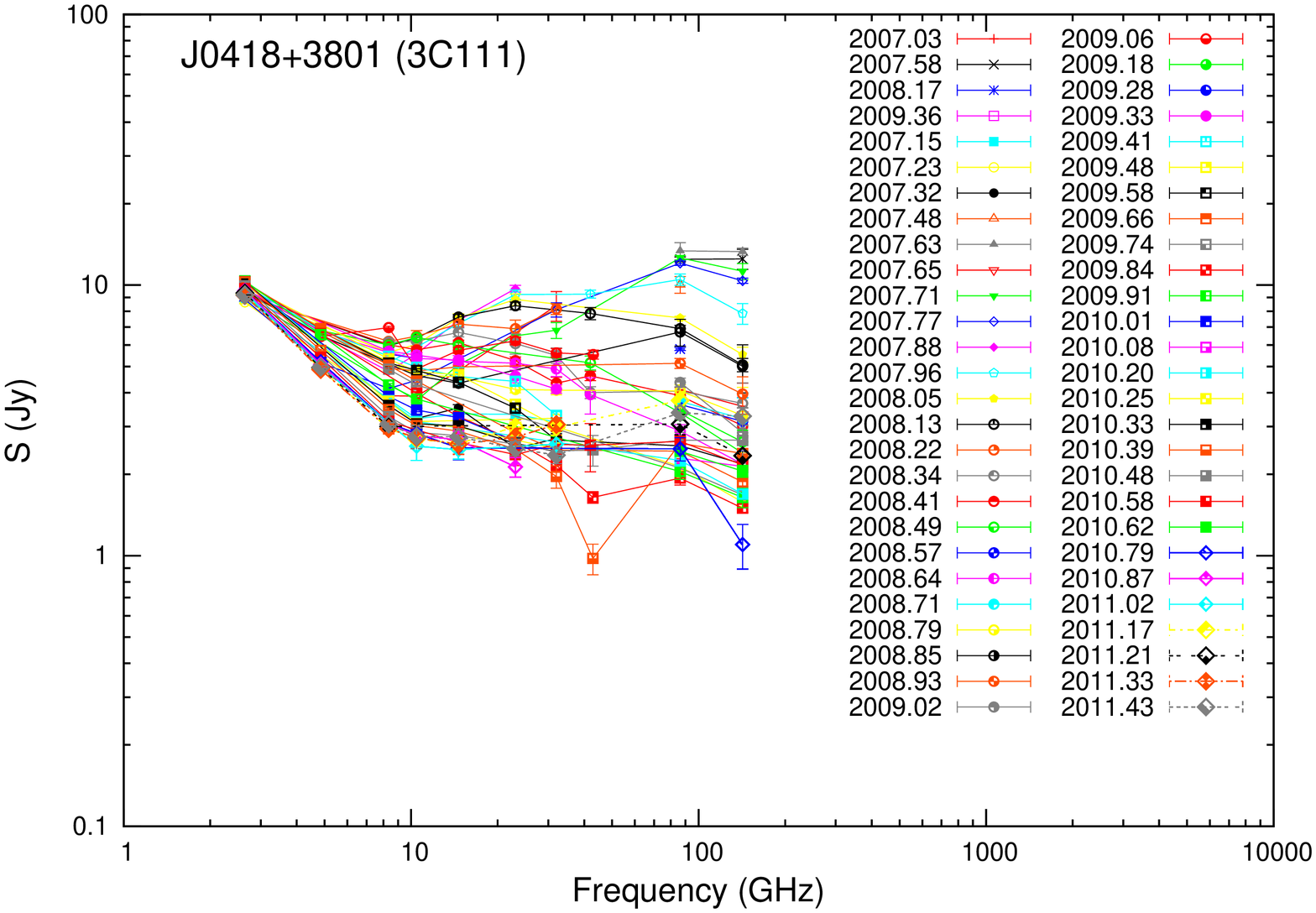} \label{fig:t3b}}
  \subfigure[Type 4]{\includegraphics[width=70pt,height=150pt,angle=-90]{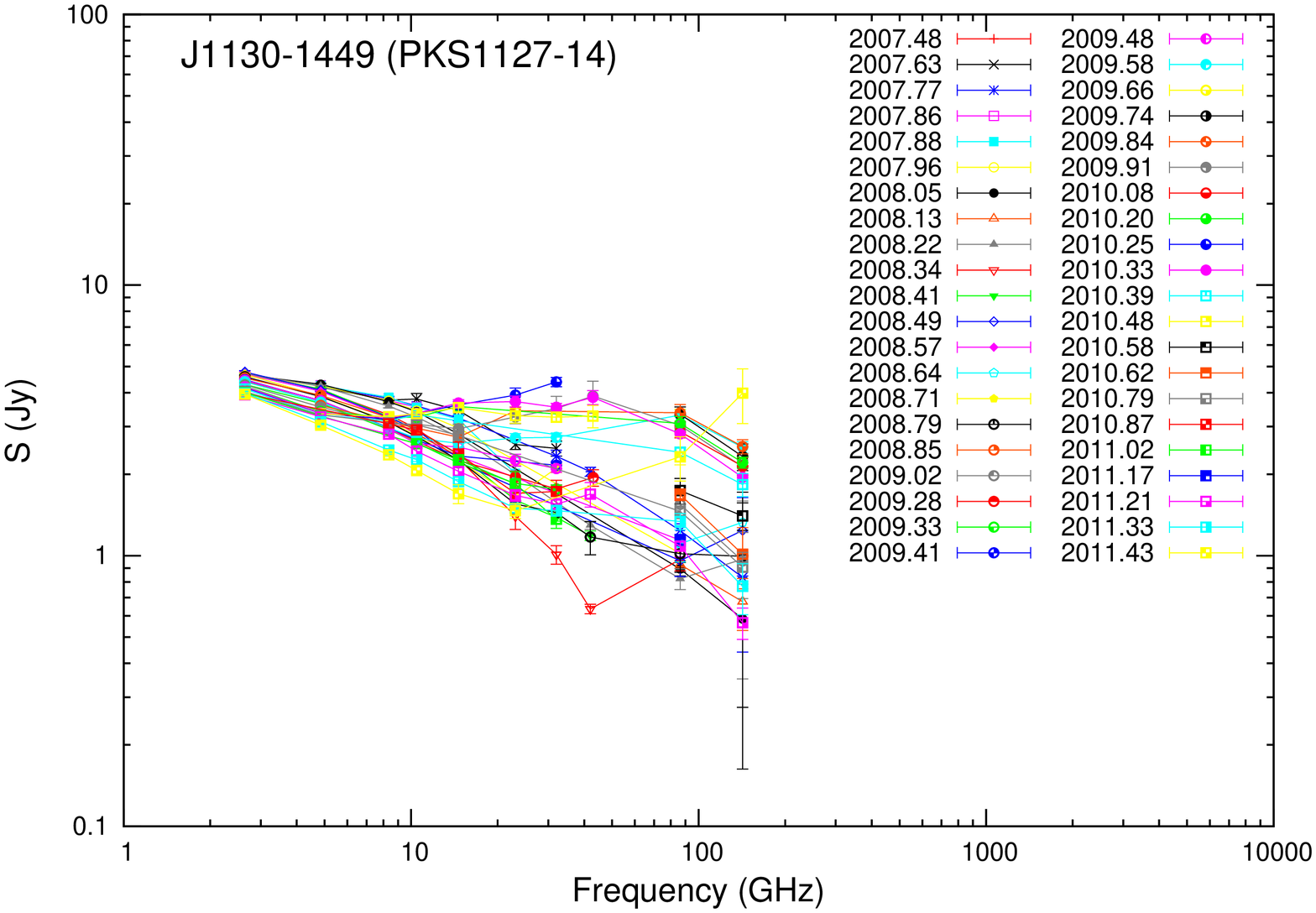} \label{fig:t4}}
  \\
  \subfigure[Type 4b]{\includegraphics[width=70pt,height=150pt,angle=-90]{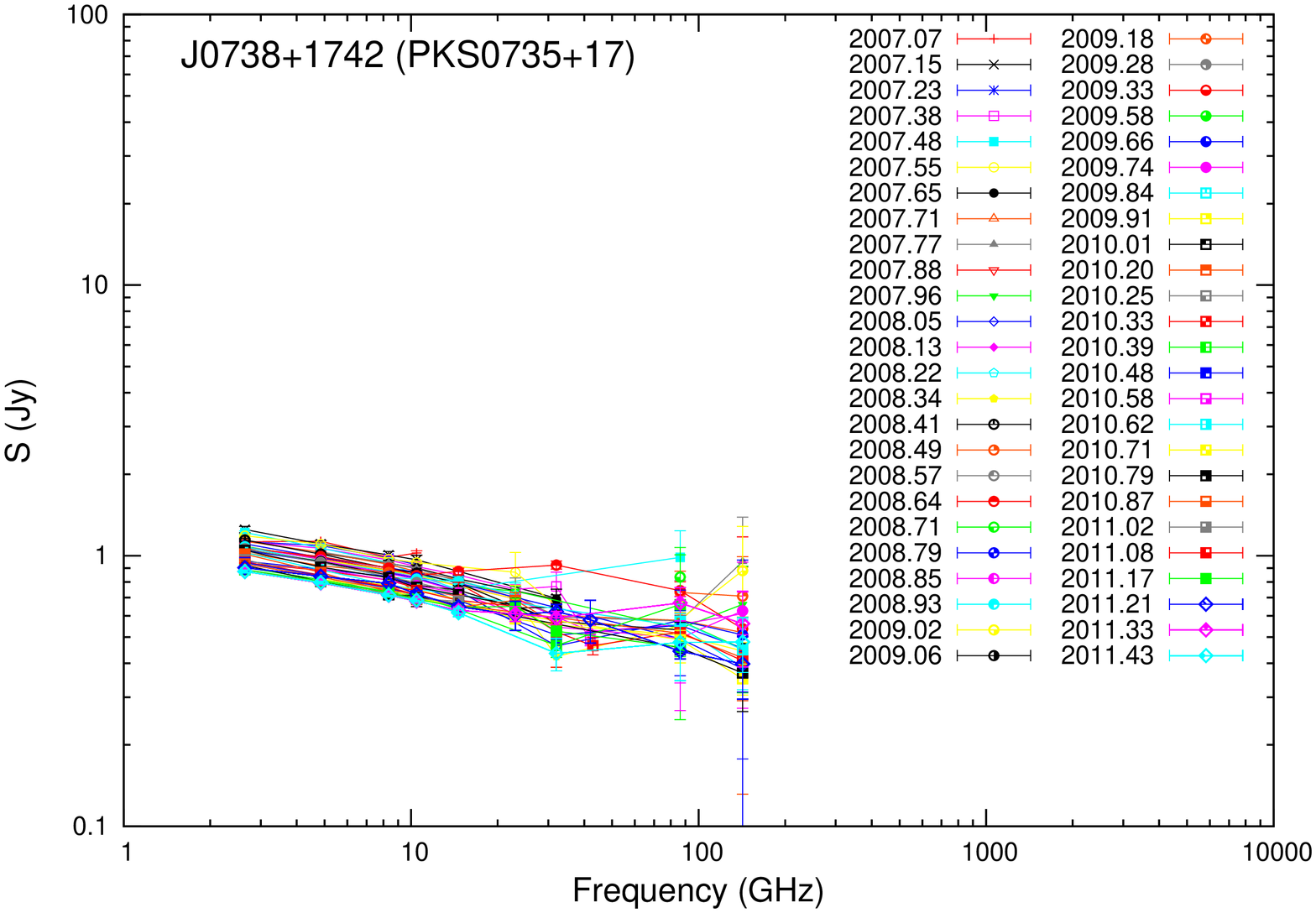} \label{fig:t4b}}  
  \subfigure[Type 5]{\includegraphics[width=70pt,height=150pt,angle=-90]{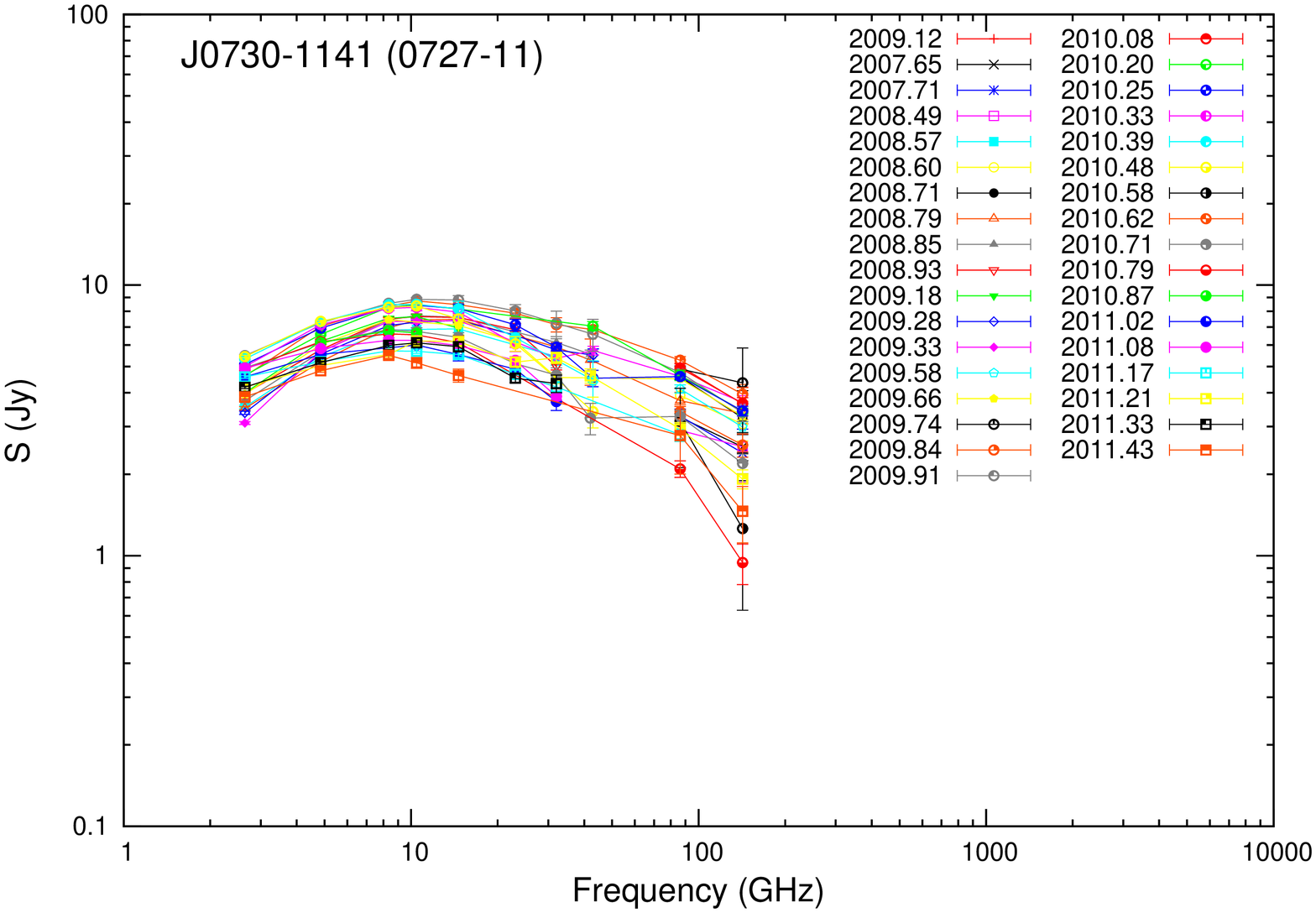} \label{fig:t5}}
  \subfigure[Type 5b]{\includegraphics[width=70pt,height=150pt,angle=-90]{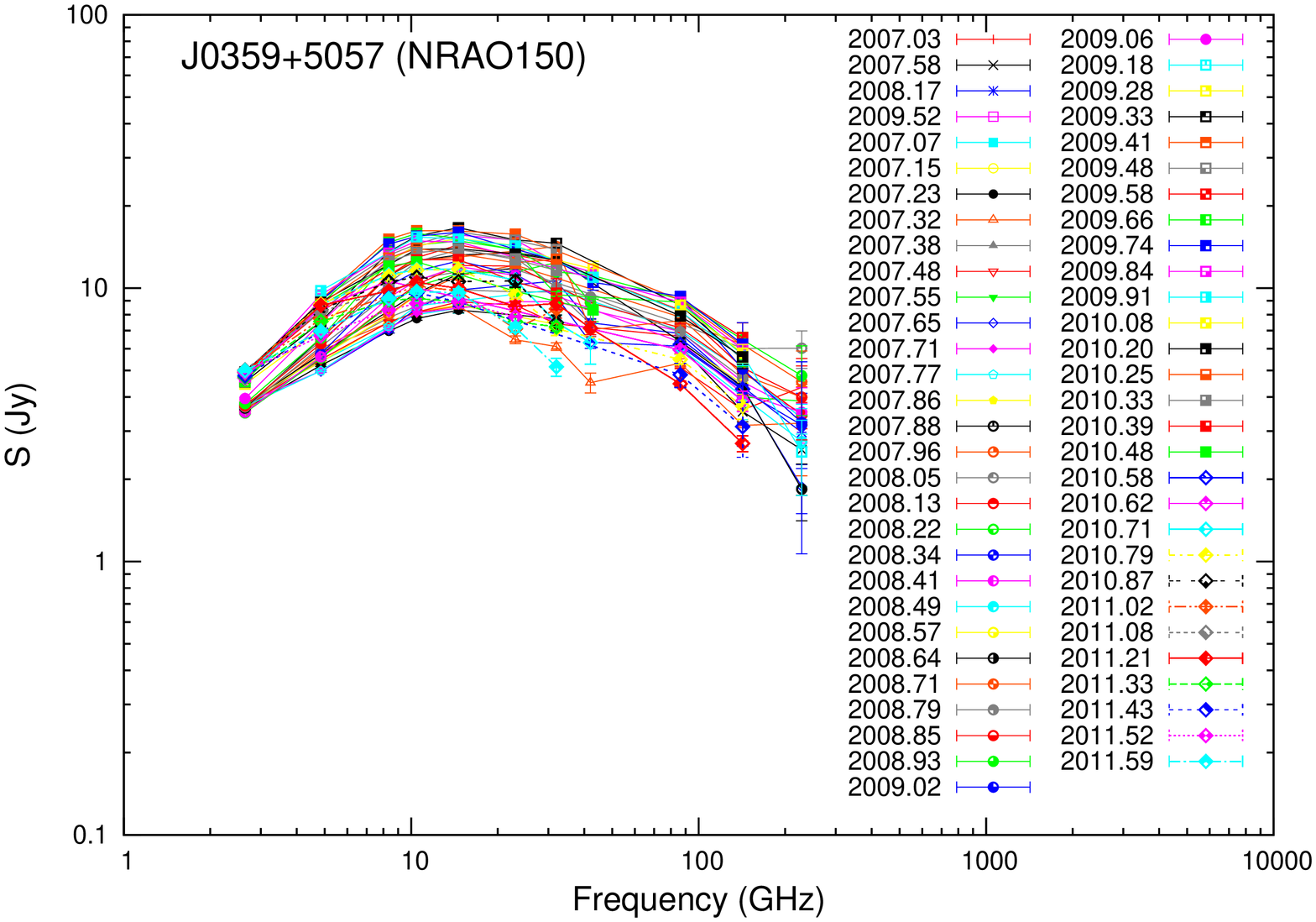} \label{fig:t5b}}
  \\
 \caption{The prototypes of the 5 types and their sub-types.}
  \label{fig:examples}
\end{figure*}
The observations discussed here have been conducted quasi-simultaneously with the
Effelsberg 100-m and the IRAM 30-m telescope (the combined spectra coherency time is a few
days) within the {\em F-GAMMA} program
\citep{fuhrmann2007AIPC,angelakis2008MmSAI..79.1042A,2010arXiv1006.5610A}. The 100-m
telescope has been observing between 2.64 and 43.05\,GHz at 8 frequencies and the 30-m
telescope at 86 and 142\,GHz (details in Fuhrmann et al. in prep., Nestoras et al. in
prep., Angelakis et al. in prep.). In the current study only data collected until June 2011, have
been used.

The data reduction includes: (a) {\em Pointing correction}, (b) {\em Elevation dependent
  gain correction}, (c) {\em Atmospheric opacity correction}, (d){\em Absolute calibration
  (sensitivity correction)}. The overall uncertainties reached for the {\em F-GAMMA}
program are of the order of 0.5 - 5\,\% for Effelsberg and of the order of $\le10$\,\% for
IRAM. More details can be found in \cite{angelakis2009AnA} as well as in Angelakis et
al. (in prep.), Fuhrmann et al. (in prep.) and Nestoras et al. (in prep.).

\section{ANALYSIS}

\subsection{Phenomenological Classification of the Variability Patterns}
\label{subsec:classification}
The visual inspection of the examined sources reveals a plurality in spectral features as
well as in the variability pattern that different sources exhibit. Despite the apparent
complexity it appears that any of the 78 sources studied here, can be classified in one
among only five phenomenological classes on the basis of its variability pattern, which
are termed numerically from 1 to 5 (more details will be given by Angelakis et al. in
prep.). Four of them show also sub-types which however do not deserve a separate type and
are named after the main type followed by the letter ``b''. The prototype sources are
shown in Figures~\ref{fig:t1}--\ref{fig:t5b}. Their phenomenological characteristics, are:
\\
{\bf Type 1}: is clearly dominated by spectral evolution. At an instant in time the
spectrum appears convex and its peak is drifting within the observing band-pass from high
towards lower frequencies, covering a significant area in the $S-\nu$ space.  The convex
component is smoothly changing towards an ultimate flat or mildly steep power-law which is
then followed by consequent events. There is no evidence for a stable steep spectrum. The
lowest frequencies in the bandpass are remarkably variable indicating that the activity
seizes at frequencies much lower than the lowest in our band-pass. The prototype source is
shown in Figure~\ref{fig:t1}.
\\
{\bf Type 1b}: As a sub-class of the previous one, type 1b shows similar characteristics
except that the lowest frequency does not show as intense variability. The activity seizes
around this part of the band-pass (see Figure~\ref{fig:t1b}).
\\
{\bf Type 2}: is also dominated by spectral evolution. The basic characteristic of
  this case is the fact that the flux density at the lowest frequency during the steepest
  spectrum phase is higher than that during the inverted spectrum phase. Moreover, the
  maximum flux density reached by the flaring events is significantly above that at the
  lowest frequency. This implies that the observed steep spectrum is not a quiescent
  spectrum but rather the ``echo'' of an older, yet recent, outburst. The prototype of
  this type is shown in Figure~\ref{fig:t2}.
  \\
  {\bf Type 3}: Type 3, shown in Figure~\ref{fig:t3}, is also dominated by spectral
  evolution. The identifying characteristics of this type are: (a) the fact that the
  lowest frequency practically does not vary and, (b) the maximum flux density level
  reached by outbursts is comparable to that at the lowest band-pass frequency. This
  phenomenology leaves hints that the events seize very close to the lowest frequency of
  the band-pass and hence a quiescent spectrum is becoming barely evident.
\\
{\bf Type 3b}: Type 3b, shown in Figure~\ref{fig:t3b}, is very similar to type 3. Here
however the quiescent spectrum is seen clearly at least at the 2 lowest frequencies.
\\
{\bf Type 4}: Sources of this type spend most of the time as steep spectrum ones which
are sometimes showing an outburst of relatively low power propagating towards low
frequencies. A representative case is shown in Figure~\ref{fig:t4}.
\\
{\bf Type 4b}: This type includes persistently steep spectrum cases as it is shown in
Figure~\ref{fig:t4b}.

All previous classes, are clearly dominated by spectral evolution. There exists a class of
sources for which the variability happens self-similarly without signs of spectral
evolution. Those are grouped in a separate type with two sub-types:
\\
{\bf Type 5}: In this case the spectrum is convex and follows an ``achromatic''
evolution. That is, it shifts its position in the $S-\nu$ space preserving its shape. This
is shown clearly in Figure~\ref{fig:t5}.
\\
{\bf Type 5b}: This type shows, in principle, characteristics similar to the previous
one but there occurs a mild yet noticeable shift of the peak ($S_\mathrm{m},\nu_\mathrm{m}$)
towards lower frequencies as the peak flux density increases. A characteristic case is
shown in  Figure~\ref{fig:t5b}.

This classification is done solely on the basis of the phenomenological characteristics of
the variability pattern shown by the radio spectra within a given band-pass. As it is
discussed in the next section, it appears that all the phenomenology for types 1--4b can
be naturally explained with the same underlying system observed under different
circumstances.

\subsection{A Physical Interpretation of the Variability Types 1 -- 4b}
\begin{figure}[]
\includegraphics[width=0.3\textwidth]{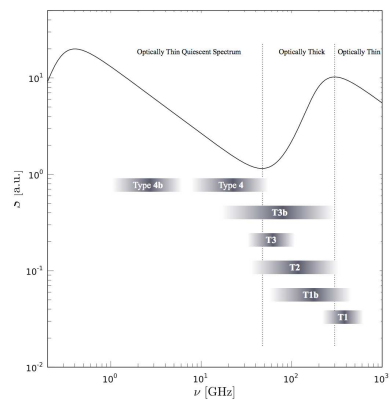}\hspace{2pc}%
\begin{minipage}[b]{14pc}\caption{\label{fig:principal}The assumed two-component
    system. The different variability types can be reproduced with the appropriate
    modulation of the relative position and relative broadness of the band-pass denoted by
    the grey shaded areas.}
\end{minipage}
\end{figure}

The phenomenological types 1 -- 4b discussed earlier can be reproduced by the same simple
two-component system, made of: (a) A power-law quiescent spectrum attributed to the
optically thin emission (large-scale jet and recent flaring events) and (b) the convex
synchrotron self-absorbed spectrum of a current outburst superimposed on the quiescent
part. The assumed configuration is presented in Figure~\ref{fig:principal} where the
shaded areas denote the observing band-pass. The phenomenology shown there captures the
system (solid line) at an instant in time and the spectral shape that would be observed
depends on: (a) the {\bf \em position} of the shaded areas relative to the high and low
frequency peak (i.e. the peak of the outburst) and (b) the {\bf \em width} of the
band-pass relative to the width of the bridge between the optically thick part of the
outburst and the steep part of the quiescent spectrum.

These two quantities can be modulated by the combination of (a) the {\bf \em redshift} and
(b) the {\bf \em source intrinsic properties}. The {\em redshift} changes the relative
{\em position} of the band-pass allowing a different part of the spectrum to be
sampled. The {\em source intrinsic properties} imply that different sources show
different spectral characteristics (e.g. peak frequency of the outburst, peak flux density
excess of the outburst over the quiescent spectrum, broadness of the valley etc.). More
importantly, the dynamical evolution of a flaring event, is a function of the {\em source
  intrinsic properties} and introduces a third parameter (c) the {\em flare specific
  properties} which is determining the characteristics of the variability pattern.

In order to examine whether the assumed model can reproduce the observed phenomenologies,
some characteristic cases have been evaluated.

\subsection{Reproducing the Observed Phenomenologies 1 -- 4b}
\begin{figure*}
  \centering
  \subfigure[ powerful source at $z=1.5$.]{\includegraphics[width=150pt,height=80pt]{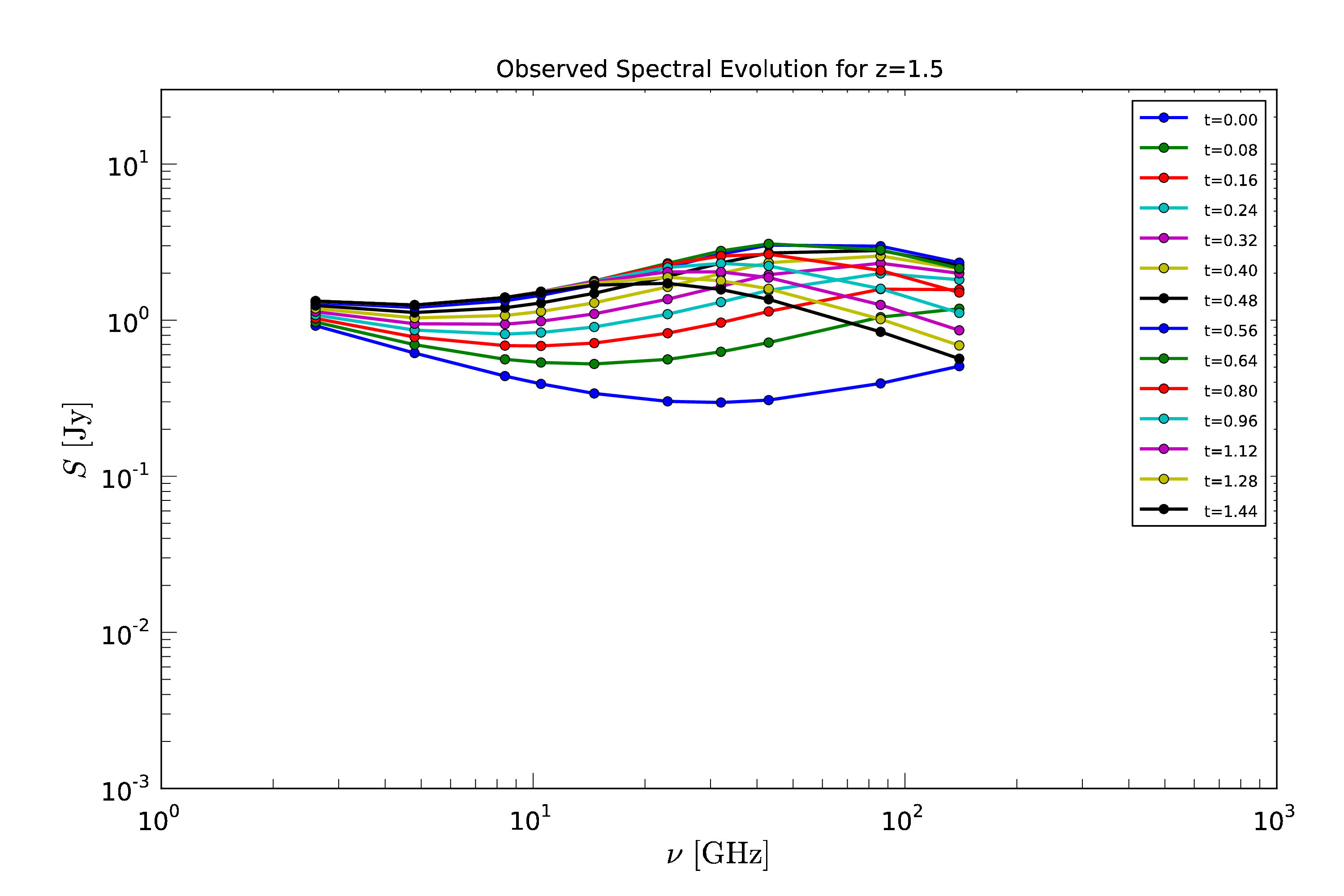} \label{fig:high15}}  
  \subfigure[ medium source at $z=1.5$.]{\includegraphics[width=150pt,height=80pt]{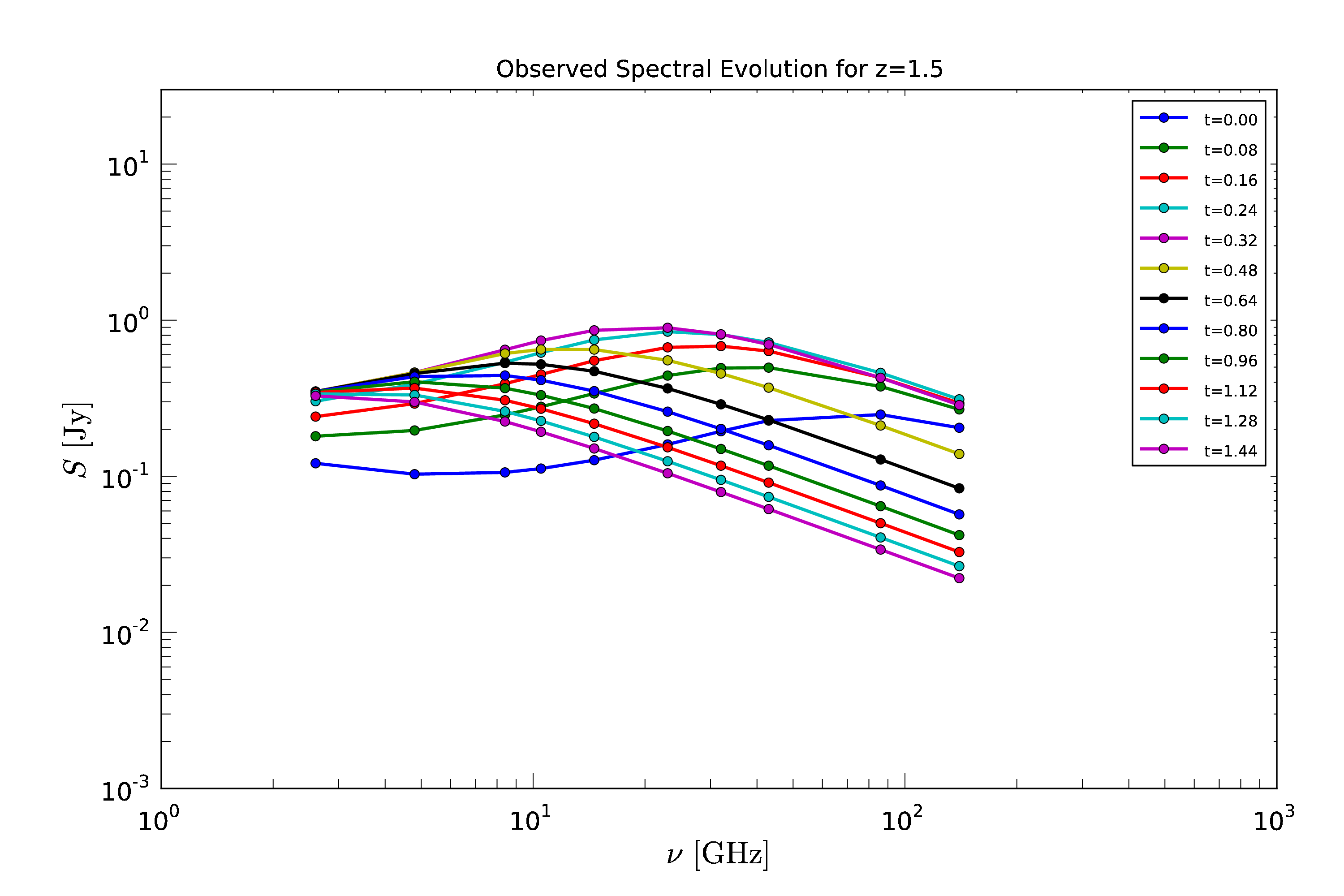} \label{fig:med15}}  
  \subfigure[ weak source at $z=1.5$.]{\includegraphics[width=150pt,height=80pt]{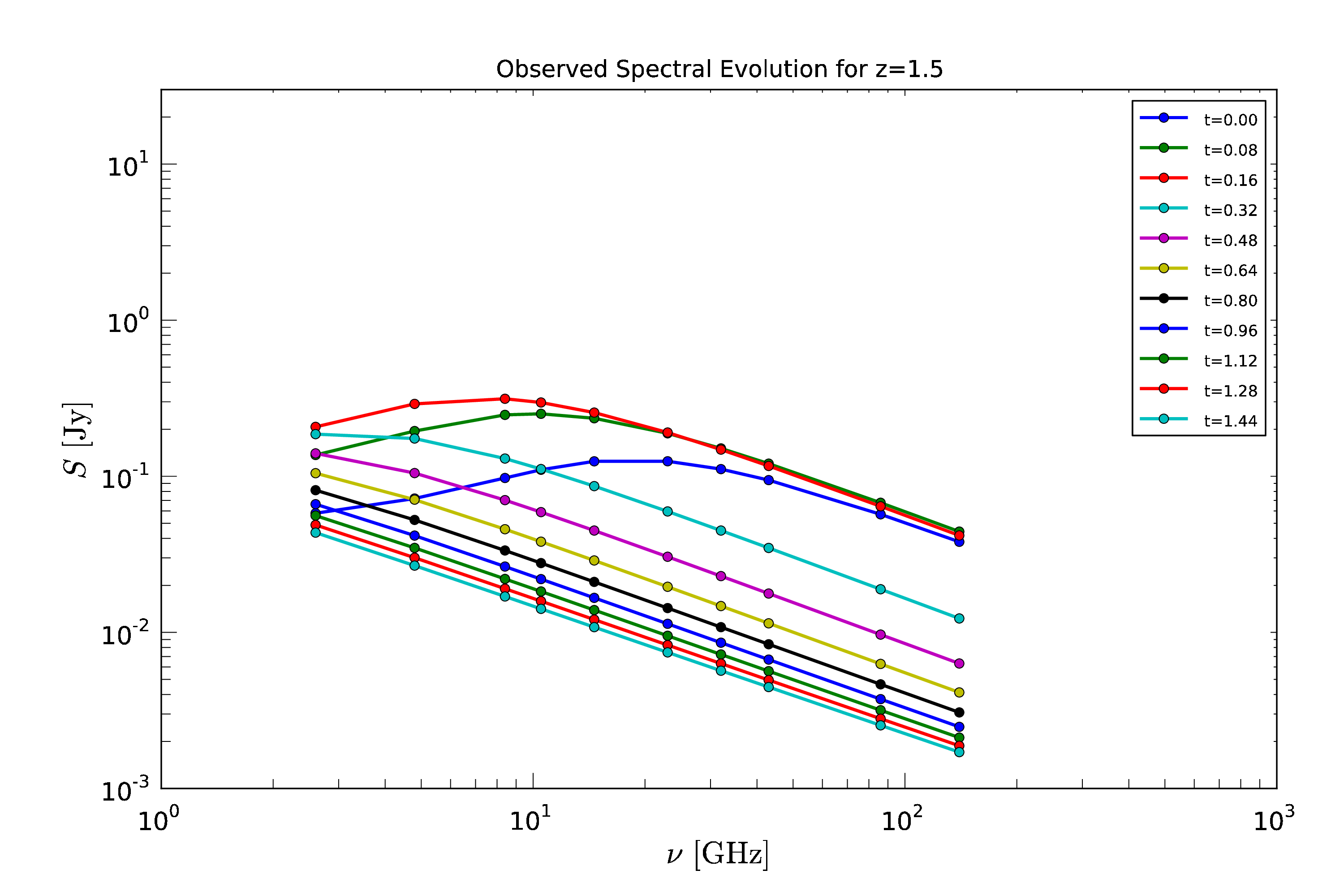} \label{fig:weak15}}  
 \\
\caption{Three characteristic cases of the system assumed for the reproduction of the
  phenomenologies of types 1--4b.}
  \label{fig:simulations}
\end{figure*}

Following the hypothesis that all the observed events are the reflection of the same
process, namely shocks evolving in jets seen with different frequency band-passes at
different evolutionary stages, the shock-in-jet model
\cite{Marscher1985ApJ,turler2000AnA...361..850T} has been applied to reproduce their
temporal evolution. The followed approach is presented in \cite{2011AnA...531A..95F} where
the flaring event passes different radiative evolutionary stages ({\sl Compton}, {\sl
  Synchrotron} and {\sl Adiabatic} stage). Simulations have been made for a large fraction
of the parameter space. In Figure~ \ref{fig:simulations} are presented only three cases of
sources  at $z=1.5$ but of different luminosity. From these plots it is already
clear that the assumed scenario can reproduce most of the observed phenomenologies. 

\section{DISCUSSION}
The variability patterns of the studied blazars can be categorized in (a) spectral
evolution dominated cases and (b) self-similarly varying convex spectra ones. This implies
that there must exist two distinct mechanisms causing variability. This refers to the
available baseline (roughly five years) meaning that sources of type 5 and 5b could still
show spectral evolution over longer time scales. An additional element that points towards
a different variability mechanism is the persistency of the spectral shape, in the case of
type 5 and the fashion of change in the case of type 5b which is not seen in cases of
clear spectral evolution.

Of the 78 examined sources, 8 show achromatic variability. The interesting characteristics
is that the cases that show a mild spectral evolution (type ``5b''), the turnover flux and
frequency $S_\mathrm{m}$ and $\nu_\mathrm{m}$, are evolving in an anti-correlated fashion
(see e.g. Figure~\ref{fig:t5b}). Apart from the fact that all of them show the clear
presence of a large scale jet even at 2\,cm (as it is shown form the MOJAVE images
\citealt{Kellermann2004ApJ}) no other peculiar property has been identified so
far. Possible mechanisms that are examined to be producing this variability include:
opacity effects, changes in the magnetic field structure, changes in the Doppler factors
and geometrical effects.

None of the studied sources has shown a switch of type over the baseline of the {\em
  F-GAMMA} program, neither between types of the same underlying mechanism (i.e. 1--4b)
nor between types with different underlying mechanism (i.e. types 1--4b and types 5,
5b). This suggests that the mechanisms producing the variability is either a fingerprint
of the source or the conditions that determine it change over longer time scales. It would
be essential to observe whether a source can switch from achromatic to an evolution
dominate type or even further whether it exhibits periods of either variability
flavor. In any case, the persistency of the evolution dominated types implies that the
power deposited in each event for a certain source is not varying significantly from one
event to the other. Further investigations to investigate this statement are underway and
will be presented elsewhere.

Concerning the evolution dominated case, it seems that the \cite{Marscher1985ApJ} model
provides a precise reproduction of the observed phenomenology and most importantly, over a
range of intrinsic parameters covered by the {\em F-GAMMA} sample. Studies to examine
whether other variability mechanisms can reproduce the observed phenomenology (shapes,
time scales etc.) are needed. In any case, any successful model could be used for
reversing the process and be used for calculating physical parameters from the observed
spectra.

\bigskip 
\begin{acknowledgments}
  Based on observations with the 100\,m telescope of the MPIfR (Max-Planck-Institut f\"ur
  Radioastronomie) and the IRAM 30-m telescope. I. Nestoras and R. Schmidt are members of
  the International Max Planck Research School (IMPRS) for Astronomy and Astrophysics at
  the Universities of Bonn and Cologne. The {\em F-GAMMA} team sincerely thanks
  Dr. A. Kraus and the Time Allocation Committee of the 100-m and 30-m telescope for
  supporting the continuation of the program. \end{acknowledgments}

\bigskip 

\end{document}